\documentclass[11pt]{article}
\usepackage{geometry,url}                % See geometry.pdf to learn the layout options. There are lots.
\usepackage{graphicx}

\usepackage{amssymb}
\usepackage{epstopdf}

\usepackage{mathptmx}               %% chooses math times font
\usepackage[scaled=.92]{helvet}  %% chooses appropriately
\if01

\fi

\newcounter{listing}
\newcommand{\Q}{{\bf Q}} 
\newcommand{\Z}{{\bf Z}}

\let\oldcal\cal \renewcommand{\cal}[1]{{\oldcal #1}}

\begin{document}

\title{De-Quantising the  Solution of Deutsch's Problem}
\author{Cristian S. Calude\\
\phantom{xx}\\
Department of Computer Science\\ University of
Auckland, New Zealand\\ \url{www.cs.auckland.ac.nz/~cristian}}  
\maketitle
%\section{}
%\subsection{}

\begin{abstract}
Probably the simplest and most frequently used way to illustrate the power of quantum computing is to solve the so-called
{\it Deutsch's problem}. Consider a Boolean function $f: \{0,1\} \rightarrow \{0,1\}$
and suppose that we have a (classical) black box to compute it.  The problem asks 
whether $f$ is constant (that is, $f(0) = f(1)$) or balanced ($f(0) \not= f(1)$).    Classically, to  solve the problem  seems to require the computation of $f(0)$ and $ f(1)$, and then  the  comparison of results.
Is it possible to solve the problem with {\em only one}  query on $f$? In a famous paper published in 1985,  Deutsch posed the problem and obtained a ``quantum''  {\em partial affirmative answer}.  In 1998 a complete, probability-one
solution was presented by Cleve, Ekert, Macchiavello, and Mosca.   Here we will show that the quantum solution can be {\it de-quantised} to a deterministic simpler solution which is as efficient as the quantum one.  The use of ``superposition'', a key ingredient of quantum algorithm, 
is---in this specific case---classically  available.
\end{abstract}
\thispagestyle{empty}

\section{Introduction}
Consider a Boolean function $f: \{0,1\} \rightarrow \{0,1\}$
and suppose that we have a black box to compute it.  
Deutsch's problem asks to test
whether $f$ is constant (that is, $f(0) = f(1)$) or balanced ($f(0) \not= f(1)$) allowing  {\em only one query} on the black box computing $f$.

Our aim is to show a simple deterministic classical solution
to
Deutsch's Problem. To be able to compare the quantum and classical solutions we will present both solutions in detail.

\section{The quantum solution}
The quantum  technique is to ``embed'' the classical  computing box (given by 
$f$) into a quantum box,  then use the quantum device on a ``superposition'' state, and finally make a single measurement of the output of the quantum computation.
This technique was proposed by Deutsch in the famous paper \cite{D85}; the problem was extended by Deutsch and Josza \cite{DJ92} and fully solved with probability one by Cleve, Ekert, Macchiavello, and Mosca \cite{CEMM} (see  Gruska \cite{Gruska}, or Nielsen and Chuang	\cite{NC}).

 Suppose that we have a quantum black box to compute $f_{Q}$ which extends $f$ from $\{0,1\}$ to the quantum (Hilbert) space generated by the base  $\{|0\rangle, |0\rangle\}$. 
 %(which you recall form the basis of $\IC^{2}$).
This means, that $f(0) = f_{Q}(|0\rangle)$ and  $f(1) = f_{Q}(|1\rangle)$.
The quantum computation of  $f_{Q}$ will be done  using
 the transformation
$U_f$  which applies to two Qbits, $|x\rangle$ and $|y\rangle$, and produces
$|x\rangle |y \oplus f(x)\rangle$ ($\oplus$ denotes the sum modulo 2).
 This transformation flips the second Qbit
 if $f$ acting on the first
Qbit is 1, and does nothing if $f$ acting on the first
Qbit is 0.

Here is a standard mathematically formulation of the  quantum algorithm. Start with $U_{f}$ and evolve it  on a superposition
of $|0\rangle$  and $|1\rangle$. Assume first that the second Qbit is initially prepared
in the state $\frac{1}{\sqrt{2}}(|0\rangle - |1\rangle )$. Then,

\begin{eqnarray*}
 U_f \left(|x\rangle  \frac{1}{\sqrt{2}}(|0\rangle - |1\rangle )\right) 
 & = & 
|x\rangle \frac{1}{\sqrt{2}}(|0\oplus f(x) \rangle - |1 \oplus f(x)\rangle )\\
&=& (-1)^{f(x)} |x\rangle \frac{1}{\sqrt{2}} (|0\rangle - |1\rangle ).\\
\end{eqnarray*}

Next take the first Qbit to be $ \frac{1}{\sqrt{2}}(|0\rangle + |1\rangle). $
The quantum black box will produce

\begin{equation}
\label{qfunction}
U_f  \left( \frac{1}{\sqrt{2}}(|0\rangle + |1\rangle) \frac{1}{\sqrt{2}} (|0\rangle - |1\rangle)\right)
%%\phantom{xxxxxxxxxxxxxxxxxxxxxxxxxxxxxxxxxxx}\\
%%&& \phantom{xxxxx}  = \frac{1}{\sqrt{2}} ((-1)^{f(0)} |0\rangle + (-1)^{f%%(1)} |1
%%\rangle)  \frac{1}{\sqrt{2}} (|0\rangle - |1\rangle)\\
%\phantom{xxxxx}  
= \frac{1}{2}(-1)^{f(0)} (|0\rangle + (-1)^{f(0) \oplus f(1)} |1\rangle)(|0\rangle - |1\rangle ).
\end{equation}

\medskip

Next  perform  a measurement that projects the first Qbit onto the 
 basis
 
$$\frac{1}{\sqrt{2}}(|0\rangle + |1\rangle), \frac{1}{\sqrt{2}}(|0\rangle - |1\rangle).$$

We will obtain $\frac{1}{\sqrt{2}}(|0\rangle - |1\rangle)$ if the function 
$f$
is balanced and $\frac{1}{\sqrt{2}}(|0\rangle + |1\rangle)$ in the opposite case.\\

 To better understand the action of  (\ref{qfunction}) we will present  $U_{f}$   in matrix form as:
 
\[U_{f}= \left(\begin{array}{cccc}
1-f(0)  & f(0) &  0 & 0 \\
f(0) & 1-f(0)  &  0 & 0 \\
0 & 0  &  1-f(1) & f(1) \\
0  & 0  &  f(1) & 1-f(1) \\
\end{array}\right)\raisebox{.5ex}.\] 

\if01

\[U_{f}= \left(\begin{array}{cccc}
(1-f(0), 0)  & (f(0),0)  &  (0,0) & (0,0) \\
(f(0), 0)  & (1-f(0),0)  &  (0,0) & (0,0) \\
(0, 0)  & (0,0)  &  (1-f(1),0) & (f(1),0) \\
(0, 0)  & (0,0)  &  (f(1),0) & (1-f(1),0) \\
\end{array}\right)\raisebox{.5ex}.\]

Indeed, the first column of the matrix is obtained from \[U_{f} |00\rangle = |0\rangle |0\oplus f(0)\rangle = |0 f(0)\rangle.\] 

There are two cases: if $f(0)=0$, then 
\[|0 f(0)\rangle =  \left(\begin{array}{c}
(1, 0)  \\
(0,0)\\
(0,0)\\
(0,0)\
\end{array}\right) = \left(\begin{array}{c}
(1-f(0), 0)  \\
(f(0),0)\\
(0,0)\\
(0,0)\
\end{array}\right),\] 
and  if $f(0)=1$, then 
\[|0 f(0)\rangle =  \left(\begin{array}{c}
(0,1)  \\
(0,0)\\
(0,0)\\
(0,0)\
\end{array}\right) = \left(\begin{array}{c}
(1-f(0), 0)  \\
(f(0),0)\\
(0,0)\\
(0,0)\
\end{array}\right).\] 
\fi

Whatever the values of $f(0)$ and $f(1)$, the matrix $U_{f}$ is unitary, %$U^{\dag}U=I_{4}$,
so $U_{f}$ is a legitimate quantum black box.
Next we are going to use the Hadamard transformation $H$ to generate a superposition of states:

\medskip

\[H= \left(\begin{array}{cccc}
\frac{1}{2}  & \phantom{-}\frac{1}{2} &  \phantom{-}\frac{1}{2} & \phantom{-}\frac{1}{2} \\[0.5ex]
\frac{1}{2} & -\frac{1}{2}  &  \phantom{-}\frac{1}{2} & -\frac{1}{2} \\[0.5ex]
\frac{1}{2}  & \phantom{-}\frac{1}{2}  &  -\frac{1}{2} & -\frac{1}{2} \\[0.5ex]
\frac{1}{2}  & -\frac{1}{2}  &  -\frac{1}{2} & \phantom{-}\frac{1}{2} \\
\end{array}\right)\raisebox{.5ex}.\] 

\if01
\[H= \left(\begin{array}{cccc}
(\frac{1}{2}, 0)  &(\frac{1}{2}, 0)  &  (\frac{1}{2}, 0) & (\frac{1}{2}, 0) \\
(\frac{1}{2}, 0)  &(-\frac{1}{2}, 0)  &  (\frac{1}{2}, 0) & (-\frac{1}{2}, 0) \\
(\frac{1}{2}, 0)  &(\frac{1}{2}, 0)  &  (-\frac{1}{2}, 0) & (-\frac{1}{2}, 0) \\
(\frac{1}{2}, 0)  &(-\frac{1}{2}, 0)  &  (-\frac{1}{2}, 0) & (\frac{1}{2}, 0) \\
\end{array}\right)\raisebox{.5ex}.\] \\
\fi

Here is the quantum algorithm solving Deutsch's problem:

\begin{center}
\begin{minipage}{15cm} {\tt 
\noindent 1. Start with a closed physical system prepared in the quantum state $|01\rangle$.
\\
2. Evolve the system according to $H$.\\
3.  Evolve the system according to $U_{f}$.\\
4.  Evolve the system according to $H$.\\
5. Measure the system. If the result is the second possible output, then
$f$ is constant;  if the result is the fourth possible output, then
$f$ is  balanced.}
\end{minipage}
\end{center}

To prove the correctness of the quantum algorithm,  we shall show that the first and third possible outputs can be obtained with probability zero, while one (and only one) of the second and the fourth  outcomes  will be obtained with probability one, and  the result solves correctly Deutsch's problem. 

To this aim  we follow step-by-step the quantum  evolution described by the above algorithm.

In Step 1 we start with a closed physical system prepared in the quantum state $|01\rangle$:
\if01
\[ V = \left(\begin{array}{c}
(0, 0)  \\
(1,0)\\
(0,0)\\
(0,0)\
\end{array}\right)\raisebox{.5ex}.\] 
\fi
\[ V = \left(\begin{array}{c}
0  \\
1\\
0\\
0\
\end{array}\right)\raisebox{.5ex}.\] 

After Step 2 the system has evolved in the state (which is independent of $f$):

\medskip

\[H V = \left(\begin{array}{cccc}
\frac{1}{2}  & \phantom{-}\frac{1}{2} &  \phantom{-}\frac{1}{2} & \phantom{-}\frac{1}{2} \\[0.5ex]
\frac{1}{2} & -\frac{1}{2}  &  \phantom{-}\frac{1}{2} & -\frac{1}{2} \\[0.5ex]
\frac{1}{2}  & \phantom{-}\frac{1}{2}  &  -\frac{1}{2} & -\frac{1}{2} \\[0.5ex]
\frac{1}{2}  & -\frac{1}{2}  &  -\frac{1}{2} & \phantom{-}\frac{1}{2} \\
\end{array}\right)  \times \left(\begin{array}{c}
0  \\
1\\
0\\
0\
\end{array}\right) = \left(\begin{array}{c}
\phantom{-}\frac{1}{2}\\[0.5ex]
-\frac{1}{2}\\[0.5ex]
\phantom{-}\frac{1}{2}\\[0.5ex]
-\frac{1}{2}\\
\end{array}\right)\raisebox{.5ex}.\]
\if01
\[H V = \left(\begin{array}{cccc}
(\frac{1}{2}, 0)  &(\frac{1}{2}, 0)  &  (\frac{1}{2}, 0) & (\frac{1}{2}, 0) \\
(\frac{1}{2}, 0)  &(-\frac{1}{2}, 0)  &  (\frac{1}{2}, 0) & (-\frac{1}{2}, 0) \\
(\frac{1}{2}, 0)  &(\frac{1}{2}, 0)  &  (-\frac{1}{2}, 0) & (-\frac{1}{2}, 0) \\
(\frac{1}{2}, 0)  &(-\frac{1}{2}, 0)  &  (-\frac{1}{2}, 0) & (\frac{1}{2}, 0) \\
\end{array}\right)  \times \left(\begin{array}{c}
(0, 0)  \\
(1,0)\\
(0,0)\\
(0,0)\
\end{array}\right) = \left(\begin{array}{c}
(\frac{1}{2}, 0)  \\
(-\frac{1}{2},0)\\
(\frac{1}{2},0)\\
(-\frac{1}{2},0)\
\end{array}\right)\raisebox{.5ex}.\]
\fi

\medskip

After Step 3 the quantum system is in the state (which {\em depends} upon $f$):

\medskip

\[U_{f}H V =  \left(\begin{array}{cccc}
1-f(0)  & f(0) &  0 & 0 \\
f(0) & 1-f(0)  &  0 & 0 \\
0 & 0  &  1-f(1) & f(1) \\
0  & 0  &  f(1) & 1-f(1) \\
\end{array}\right)  \times \left(\begin{array}{c}
\phantom{-}\frac{1}{2}\\[0.5ex]
-\frac{1}{2}\\[0.5ex]
\phantom{-}\frac{1}{2}\\[0.5ex]
-\frac{1}{2}\\
\end{array}\right) 
%\]
%\[\phantom{xxxxx}
 =  \left(\begin{array}{c}
\phantom{-}\frac{1}{2} - f(0)  \\[0.5ex]
-\frac{1}{2}+f(0)\\[0.5ex]
\phantom{-}\frac{1}{2}-f(1)\\[0.5ex]
-\frac{1}{2}+f(1)\\
\end{array}\right) \raisebox{.5ex}.  \]
\if01
\[U_{f}H V =  \left(\begin{array}{cccc}
(1-f(0), 0)  & (f(0),0)  &  (0,0) & (0,0) \\
(f(0), 0)  & (1-f(0),0)  &  (0,0) & (0,0) \\
(0, 0)  & (0,0)  &  (1-f(1),0) & (f(1),0) \\
(0, 0)  & (0,0)  &  (f(1),0) & (1-f(1),0) \\
\end{array}\right)  \times \left(\begin{array}{c}
(\frac{1}{2}, 0)  \\
(-\frac{1}{2},0)\\
(\frac{1}{2},0)\\
(-\frac{1}{2},0)\\
\end{array}\right) \]
\[\phantom{xxxxx} =  \left(\begin{array}{c}
(\frac{1}{2} - f(0), 0)  \\
(-\frac{1}{2}+f(0),0)\\
(\frac{1}{2}-f(1),0)\\
(-\frac{1}{2}+f(1),0)\
\end{array}\right) \raisebox{.5ex}.  \]
\fi
\medskip

After Step 4, the quantum  state of the system has become:

\[H U_{f}H V = \left(\begin{array}{cccc}
\frac{1}{2}  & \phantom{-}\frac{1}{2} &  \phantom{-}\frac{1}{2} & \phantom{-}\frac{1}{2} \\[0.5ex]
\frac{1}{2} & -\frac{1}{2}  &  \phantom{-}\frac{1}{2} & -\frac{1}{2} \\[0.5ex]
\frac{1}{2}  & \phantom{-}\frac{1}{2}  &  -\frac{1}{2} & -\frac{1}{2} \\[0.5ex]
\frac{1}{2}  & -\frac{1}{2}  &  -\frac{1}{2} & \phantom{-}\frac{1}{2} \\
\end{array}\right)  \times  \left(\begin{array}{c}
\phantom{-}\frac{1}{2} - f(0)  \\[0.5ex]
-\frac{1}{2}+f(0)\\[0.5ex]
\phantom{-}\frac{1}{2}-f(1)\\[0.5ex]
-\frac{1}{2}+f(1)\\
\end{array}\right)
%\]
%\[ \phantom{xxxxxxxxxxxxx} 
= \left(\begin{array}{c}
0 \\
1-f(0)-f(1)\\
0\\
f(1)-f(0)\\
\end{array}\right) \raisebox{.5ex}. \] 
\if01
\[H U_{f}H V = \left(\begin{array}{cccc}
(\frac{1}{2}, 0)  &(\frac{1}{2}, 0)  &  (\frac{1}{2}, 0) & (\frac{1}{2}, 0) \\
(\frac{1}{2}, 0)  &(-\frac{1}{2}, 0)  &  (\frac{1}{2}, 0) & (-\frac{1}{2}, 0) \\
(\frac{1}{2}, 0)  &(\frac{1}{2}, 0)  &  (-\frac{1}{2}, 0) & (-\frac{1}{2}, 0) \\
(\frac{1}{2}, 0)  &(-\frac{1}{2}, 0)  &  (-\frac{1}{2}, 0) & (\frac{1}{2}, 0) \\
\end{array}\right)  \times  \left(\begin{array}{c}
(\frac{1}{2} - f(0), 0)  \\
(-\frac{1}{2}+f(0),0)\\
(\frac{1}{2}-f(1),0)\\
(-\frac{1}{2}+f(1),0)\\
\end{array}\right)\]
\[ \phantom{xxxxxxxxxxxxx} = \left(\begin{array}{c}
(0, 0)  \\
(1-f(0)-f(1),0)\\
(0,0)\\
(f(1)-f(0),0)\
\end{array}\right) \raisebox{.5ex}. \] 
\fi
\medskip

Finally, in Step 5 we {\it measure} the current state of the system, that is, 
the state $H U_{f}H V$, and we get:

\begin{enumerate}
\item output 1 with probability $p_{1} =0,$
\item output 2 with probability $p_{2} = (1-f_{Q}(|0\rangle)-f_{Q}(|1\rangle))^{2}$,
\item output 3 with probability $p_{3}=0,$
\item output  4 with probability  $p_{4} = (f_{Q}(|1\rangle)-f_{Q}(|0\rangle))^{2}$.
\end{enumerate}

To conclude:

\begin{itemize}
\item if $f_{Q}(|0\rangle) = f_{Q}(|1\rangle),$  then  $f(0)+f(1) = 0$ (mod 2), $f(1)-f(0)=0;$
consequently, $p_{2} =1, p_{4}=0$.

\item if $f_{Q}(|0\rangle) \not= f_{Q}(|1\rangle),$ then  $f(0)+f(1) = 1$, $f(1)-f(0)=-1$ or $f(1)-f(0)=1$;
consequently, $p_{2} =0, p_{4}=1$.

\item the outputs 1 and 3 have each probability zero.
\end{itemize}

\if01
A compact mathematically formulation of the above quantum algorithm is the following: Start with $U_{f}$ and evolve it  on a superposition
of $|0\rangle$  and $|1\rangle$. Assume first that the second Qbit is initially prepared
in the state $\frac{1}{\sqrt{2}}(|0\rangle - |1\rangle )$. Then,

\begin{eqnarray*}
 U_f \left(|x\rangle  \frac{1}{\sqrt{2}}(|0\rangle - |1\rangle )\right) 
 & = & 
|x\rangle \frac{1}{\sqrt{2}}(|0\oplus f(x) \rangle - |1 \oplus f(x)\rangle )\\
&=& (-1)^{f(x)} |x\rangle \frac{1}{\sqrt{2}} (|0\rangle - |1\rangle ).\\
\end{eqnarray*}

Next take the first Qbit to be $ \frac{1}{\sqrt{2}}(|0\rangle + |1\rangle). $
The quantum black box will produce

\begin{eqnarray*}
&& U_f  \left( \frac{1}{\sqrt{2}}(|0\rangle + |1\rangle) \frac{1}{\sqrt{2}} (|0\rangle - |1\rangle)\right)\\
%%\phantom{xxxxxxxxxxxxxxxxxxxxxxxxxxxxxxxxxxx}\\
&& \phantom{xxxxx}  = \frac{1}{\sqrt{2}} ((-1)^{f(0)} |0\rangle + (-1)^{f(1)} |1\rangle)  \frac{1}{\sqrt{2}}
(|0\rangle - |1\rangle)\\
&&\phantom{xxxxx}  = \frac{1}{2}(-1)^{f(0)} (|0\rangle + (-1)^{f(0) \oplus f(1)} |1\rangle)(|0\rangle - |1\rangle ).
\end{eqnarray*}

\medskip

Next will perform  a measurement that projects the first Qbit onto the 
 basis
 
$$\frac{1}{\sqrt{2}}(|0\rangle + |1\rangle), \frac{1}{\sqrt{2}}(|0\rangle - |1\rangle).$$

We will obtain $\frac{1}{\sqrt{2}}(|0\rangle - |1\rangle)$ if the function 
$f$
is balanced and $\frac{1}{\sqrt{2}}(|0\rangle + |1\rangle)$ in the opposite case.\\
\fi

Deutsch's problem 
was solved with only one use of $U_f$. The solution is probabilistic, and the result is obtained with probability one. Its success relies on the following three facts: 

\begin{itemize}
\item the ``embedding'' of $f$ into $f_{Q}$ (see also the discussion in Mermin \cite{Mermin}, end of section C, p. 11),
\item  the
ability of the quantum computer to be in a superposition of states: we can check whether $f_{Q}(|0\rangle)$ is equal or not to $f_{Q}(|1\rangle)$ not by  computing $f_{Q}$ on $|0\rangle)$ and $|1\rangle$,
but on a superposition of  $|0\rangle)$ and $|1\rangle$, and
\item the possibility to extract the required information with just one measurement.
\end{itemize}

\section{De-quantising the quantum algorithm for Deutsch's problem}

We de-quantise Deutsch's algorithm in the following way. We  consider $\Q$ the set of rationals, and the  space
$\Q [i] = \{a + b i\mid a,b \in \Q\}, (i = \sqrt{-1})$. We embed the original function $f$ in $\Q [i]$ and we define the classical analogue $C_{f}$ of the   quantum evolution $U_{f}$
 acting from $\Q [i]$ to itself  as follows (compare with the formula (\ref{qfunction})):

\begin{equation}
\label{graphembed}
C_{f} (a + bi) = (-1)^{0\oplus f(0)} a + (-1)^{1 \oplus f(1)} bi.
\end{equation}

The  four different possible bit-functions $f$ induce the following four functions $C_{f}$  from $\Q [i]  $  to $\Q [i] $ ($\bar{x}$ is the conjugate of $x$):%\\[-5ex]
\begin{eqnarray*}
C_{00}(x)   & =  & \bar{x}, \mbox{ if }  f(0)=0, f(1)=0,\\
C_{01}(x)  & =  & x,    \mbox{ if }  f(0)=0, f(1)=1,\\
C_{10}(x) & =  & -x,  \mbox{ if } f(0)=1, f(1)=0,\\
C_{11}(x) & =  &  -\bar{x},   \mbox{ if }  f(0)=1, f(1)=1.
\end{eqnarray*}

Deutsch's problem becomes the following:
\begin{quote} {\it A function $f$ is chosen from the set  $\{C_{00}, C_{01}, C_{10}, C_{11}\}$ and  the problem is to determine, with a single
query,  which type of function it is, balanced or constant.}
\end{quote}

%\medskip

The following {\it deterministic} classical algorithm solves the problem: 
\begin{center}
\begin{minipage}{15cm}
{\tt  Given $f$, calculate $(i-1)\times f(1+i)$. If the result is real, then the function  is
balanced; otherwise, the function is constant.}
\end{minipage}
\end{center}

Indeed, the algorithm is correct because if we calculate $(i-1)\times f(1+i)$ we get:

\medskip

\noindent \phantom{xxxxxxxxxxx} $(i-1)\times  C_{00}(1+i) = (i-1) (1-i) = 2i$,\\
\phantom{xxxxxxxxxxx} $(i-1)\times  C_{01}(1+i) = (i-1) (1+i) = -2$,\\
\phantom{xxxxxxxxxxx} $(i-1) \times  C_{10}(1+i) = (i-1) (-1-i) = 2$,\\
\phantom{xxxxxxxxxxx} $(i-1) \times C_{11}(1+i) = (i-1) (i-1) = -2i$.

\medskip

If the answer is real, then the function is balanced, and if the answer
is imaginary, then the function is constant.

\medskip

Actually, there are infinitely many similar solutions, namely for every rational $a \not=0$:

\begin{center}
\begin{minipage}{15cm}
{\tt  Given $f$, calculate $a(i-1)\times f(1+i)$. If the result is real, then the function  is
balanced; otherwise, the function is constant.}
\end{minipage}
\end{center}

\begin{center}
\begin{minipage}{15cm}
{\tt  Given $f$, calculate $a(i+1)\times f(1+i)$. If the result is real, then the function  is
constant; otherwise, the function is balanced.}
\end{minipage}
\end{center}

Of course, $\Q[i]$ plays no special role ``by itself'' in the above solution. The explanation is not  deep, just the fact that classical bits are one-dimensional
while complex numbers  are two-dimensional.  Thus one can have
``superpositions" of different basis vectors. 

\medskip

Two-dimensionality can be obtained in various other simpler ways. For example, we can choose as space the set $\Z[\sqrt{2}] = \{a+b \sqrt{2} \mid a,b \in \Z\}$, where $\overline{a+b \sqrt{2}} = a - b \sqrt{2}$. Using a similar embedding function as (\ref{graphembed}), 
$C_{f} (a + b \sqrt{2}) = (-1)^{0\oplus f(0)} a + (-1)^{1 \oplus f(1)} b \sqrt{2},$ now acting 
on $\Z[\sqrt{2}]$, we get the solution:\footnote{In fact we don't need the whole set $\Z[\sqrt{2}]$, but its finite subset $ \{a+b \sqrt{2} \mid a,b \in \Z, |a|, |b|\le 3\}$.}
\begin{center}
\begin{minipage}{15cm}
{\tt  Given $f$, calculate $(\sqrt{2}-1)\times f(1+\sqrt{2})$. If the result is rational, then the  function  is
balanced; otherwise, the function is constant.}
\end{minipage}
\end{center}

The correctness follows from the simple calculation of
$(\sqrt{2}-1)\times  f(1+\sqrt{2})$:

\medskip

 \noindent \phantom{xxxxxxxxxxxx} $(\sqrt{2}-1) \times  C_{00}(1+\sqrt{2}) = (\sqrt{2}-1) (1-\sqrt{2}) = 2 \sqrt{2}-3$,\\
\phantom{xxxxxxxxxxxx} $(\sqrt{2}-1) \times C_{01}(1+\sqrt{2}) = (\sqrt{2}-1) (1+\sqrt{2}) = 1$,\\
\phantom{xxxxxxxxxxxx} $(\sqrt{2}-1) \times C_{10}(1+\sqrt{2}) = (\sqrt{2}-1) (-1-\sqrt{2}) = -1$,\\
 \phantom{xxxxxxxxxxxx} $(\sqrt{2}-1) \times C_{11}(1+\sqrt{2}) = (\sqrt{2}-1) (\sqrt{2}-1) = 3-2\sqrt{2}$.

\medskip

If the answer is rational, then the function is balanced, and if the answer
is irrational, then the function is constant. We can classically distinguish between  1 and $ 3-2\sqrt{2}$ because $\sqrt{2}$ is computable.

So, again, the technique of ``embedding'' and ``superposition'' produces the desired result; this time, the computation is not only classical and  simpler,  but
also deterministic. 

\section{Conclusion}

We have shown a classical simple way to de-quantise the quantum solution for the Deutsch's problem. The same quantum technique,
embedding plus computation on a ``superposition'', leads to a  classical solution which is as efficient as the quantum one. More, the quantum solution is probabilistic, while the classical solution is deterministic.

How does the classical solution compare with the quantum one in terms of physical resources?  A simple analogical scheme  can implement the classical solution with two
registers each using a real number as in the quantum case when we need just two Qbits. However, a more realistic analysis should involve the complexity of the black box, the complexity of the implementation of the embedding, as well as the complexity of the query performed.

\if01
The whole point of Deutsch's algorithm was to distinguish the four
binary functions from one bit to one bit using a single query.    Deutsch's algorithm was important because
it wasn't clear until then that quantum Boolean gates offer any advantage over classical Boolean gates. However,  there are equally powerful classes of classical functions, like the one  described in this note.
\fi

The  downside  is that the superposition doesn't scale with the
idea below. It is not difficult to obtain a similar solution  for fixed $n$, but not uniformly (in each case a different function is used).
%it takes exponentially long to read in the general state
%with $n$ basis vectors, since you have $n$ choose $k$ monomials  %of degree $k$. 
Of course, uniformly the solution discussed in this note is not scalable, because  $n$ Qbits can represent $2^n$
states at the same time, which outgrows any linear function of $n$ (see \cite{DJ92}).
%But this is no reason for not having a non-uniform scalability.
%, at least for reasonable small values of $n$. 

Due to the fact that the number of efficient quantum
algorithms is still extremely small, one can speculate that, in practice,   ``hybrid-like''
algorithms may be preferable than pure quantum algorithms.

\section*{Acknowledgement}
I am indebted to Mike Stay for illuminating discussions and criticism which contributed  essentially to this note. I  thank  Vladimir Buzek, Jozef Gruska, Rossella Lupacchini, and Karl Svozil for various useful comments, specifically regarding   ``implementation'' as a  way to compare the
quantum and classical solutions.

\end{document}